# Numerical Analysis of Thermo-Hydraulic Performance of FLiBe Flowing Through a Louvered-Fin Compact Heat Exchanger

Bahadır Doğan[1], M. Mete Ozturk[2], L. Berrin Erbay[1, *]

[1] Department of Mechanical Engineering, School of Engineering and Architecture, Eskişehir Osmangazi University, Batı-Meşelik 26480, Eskişehir, Turkey

[2] Transportation Vocational School, Eskişehir Technical University, 26140 Eskişehir, Turkey

**\* Corresponding Author:** L. Berrin ERBAY, Prof. Dr.

e-mail: lberbay@ogu.edu.tr

Address: Department of Mechanical Engineering, School of Engineering and Architecture, Eskişehir Osmangazi University, Batı-Meşelik 26480, Eskişehir, Turkey

Tel: +90 222 224 1391
Fax: +90 222 224 1392

**Word Count:** 8300 (without figure and table captions)
**Figure Count:** 24 (with sub-figures)
**Table Count:** 1
**Similarity Index:** 12% (Ithenticate report without keywords, nomenclature and reference list)

# Numerical Analysis of Thermo-Hydraulic Performance of FLiBe Flowing Through a Louvered-Fin Compact Heat Exchanger

Bahadır Doğan[1], M. Mete Ozturk[2], L. Berrin Erbay[1, *]

[1] Department of Mechanical Engineering, School of Engineering and Architecture, Eskişehir Osmangazi University, Batı-Meşelik 26480, Eskişehir, Turkey

[2] Transportation Vocational School, Eskişehir Technical University, 26140 Eskişehir, Turkey

**Abstract:** The study focuses on the steady-state thermohydraulic behavior of the LiF-BeF2 (FLiBe) molten salt flowing through a compact heat exchanger with multi louvered fins in the cooling loop of a Molten Salt Reactor (MSR). This numerical benchmark study was conducted to obtain data both for baseline and cases with slow coolant circulation problems in the reactor. A series of numerical tests are conducted for low Reynolds numbers of 100 – 500 based on the louver pitch with different fin pitches (1.5, 2.0, and 2.5 mm) and fin louver angles (20º, 24º, 28º, 32º, 36º). The steady-state velocity profiles, temperature, and pressure fields are studied in detail. The heat transfer coefficient and pressure drop data are reported in terms of Colburn j-factor and Fanning friction factor f, and the volume goodness factor ($j/f^{1/3}$) to get a performance comparison for FLiBe as the coolant in a Molten Salt Reactor. As regards findings of the analyses, when the louver angle steeps, the molten salt flow becomes more louvered direct which improves the interaction and heat transfer between the fin and molten salt. On the contrary, when the fin pitches increase, the flow becomes duct-directed that reducing the contact between fin and FLiBe which leads to a decline in thermal performance.

* Corresponding Author: lberbay@ogu.edu.tr (L. Berrin ERBAY)
  Department of Mechanical Engineering, School of Engineering and Architecture,
  Eskişehir Osmangazi University, Batı-Meşelik 26480, Eskişehir, Turkey

**Nomenclature**

| | |
|---|---|
| $A_a$ | air side heat transfer area, m$^2$ |
| $A_c$ | minimum flow area, m$^2$ |
| $A_{fr}$ | frontal area, m$^2$ |
| $c_p$ | specific heat at constant pressure, Jkg$^{-1}$K$^{-1}$ |
| $f$ | friction factor |
| $F_d$ | flow depth, mm |
| $F_h$ | fin height, mm |
| $F_p$ | fin pitch, mm |
| $h$ | heat transfer coefficient, Wm$^{-2}$K$^{-1}$ |
| $j$ | Colburn $j$-factor |
| $k$ | thermal conductivity, Wm$^{-1}$K$^{-1}$ |
| $L_\alpha$ | louver angle, $^\circ$ |
| $L_h$ | louver height, mm |
| $L_p$ | louver pitch, mm |
| $\dot{m}$ | mass flow rate, kg s$^{-1}$ |
| $P$ | pressure, Pa |
| $Pr$ | Prandtl number |
| $\dot{Q}$ | heat transfer rate, W |
| $Re_{Lp}$ | Reynolds number based on louver pitch |
| $St$ | Stanton number |
| $t$ | thickness of the louvered fin, mm |
| $T$ | temperature, K |
| $T_i$ | inlet temperature, K |
| $T_d$ | tube depth, mm |
| $T_p$ | tube pitch, mm |
| $T_o$ | outlet temperature, K |
| $T_s$ | surface temperature, K |
| $\Delta T_m$ | logarithmic mean temperature difference, K |
| $V$ | velocity, m s$^{-1}$ |
| $u_c$ | maximum velocity at minimum flow area, m s$^{-1}$ |
| $u_{in}$ | free flow velocity at the inlet, m s$^{-1}$ |

*Greek letters*

| | |
|---|---|
| $\alpha$ | thermal diffusivity, m$^2$s$^{-1}$ |
| $\mu$ | dynamic viscosity, kg m$^{-1}$s$^{-1}$ |
| $\rho$ | density, kg m$^{-3}$ |
| $\upsilon$ | kinematic viscosity, m$^2$s$^{-1}$ |

*Abbreviations*

| | |
|---|---|
| ANP | Aircraft Nuclear Propulsion |
| ARE | Aircraft reactor experiment |
| FOM | Figure of Merit |
| MSBR | Molten Salt Breeder Reactor |
| MSRE | Molten Salt Reactor Experiment |
| MSRs | Molten Salt Reactors |
| NGNP | Next Generation Nuclear Plant |
| ORNL | Oak Ridge National Laboratory |
| TEMA | Tubular Exchanger Manufacturers Association |
| THORIMS – NES | Thorium Molten Salt Nuclear Energy Synergetic |

## 1. Introduction

Following the nuclear accidents, public opposition, and serious climate change problem due to hazardous carbon emissions, the Next Generation Nuclear Plant (NGNP) project and IV. Generation Nuclear Reactor Systems (Gen IV) considered the Molten Salt Reactors (MSRs) as one of the six attractive promising candidates for reliable and sustainable nuclear power generation with the least carbon emission in the future.

The molten salt nuclear reactors are not the latest concern in the field of Nuclear Engineering. The initial studies on this particular topic have begun in the 1940s in Tenessee [1]. In the meantime, regarding the application of the MSRs in different fields, Aircraft Nuclear Propulsion (ANP), aircraft reactor experiment (ARE), and aircraft reactor test (ART) can be counted as the successful studies conducted by the researchers in varying projects. Following ORNL studies, Small Molten Salt Reactors (MSRs) named Fuji have been proposed and studied by Furukawa et al. [2-5] in the Thorium Molten Salt Nuclear Energy Synergetic (THORIMS-NES) system. One of the most appropriate fluids in Fuji-MSRs is FLiBe which is used for a fluid carrying the fuel as well as coolant through the heat exchangers between the core side and steam turbine side.

In the recent studies by newly constructed MSR firms, there have been different molten salt reactor designs unfortunately including no details about heat exchangers indeed. The heat exchanger design for an MSR is a challenging subject. In the historical studies, ORNL conducted several studies on the design of the heat exchangers for MSRs and put forward the key design considerations especially for shell-and-tube type heat exchangers which were well documented [6-9]. The type and the design of the heat exchangers were determined by considering the technology and standards of the shell and tube type heat exchangers. Erbay [10] has proposed and examined a combination of MSRs with a Stirling heat engine. The study indicated the dependence on the need to design a high-performance heat exchanger as the hot-end head of the Stirling with molten salt FLiBe flowing over the channel and helium inside the channels. Knowhow of heat exchangers used in MSRs has been provided by Erbay [11] to throw light on Fuji-MSRs [3] heat removal system. Though new HX designs are suggested during some projects like EVOL and SAMOFAR [12-14], they have neither been finalized nor manufactured so far.

The performance of liquid coolants is excellent for gas cooling systems to decrease the temperature drops across heat exchangers and pumping power. Molten salts are used in MSRs as fuel and coolant fluids due to their thermal advantages with outstanding characteristics [15, 16]. The molten salt coolants used in MSRs are chemically and radiologically stable at high temperatures, have low vapor pressures, whereas high boiling temperature, high specific heat, and high thermal conductivity. The good compatibility with high-temperature materials and being inexpensive are also important properties of molten salt coolants. Molten salt coolants volumetric heat capacity is almost 25% higher than pressurized water, and it is approximately five times higher than liquid sodium. Therefore, the high volumetric heat capacity and safe working pressures make compact heat exchanger design possible.

There are some difficulties in the molten salt experiments. The advantage of a high melting point constitutes a difficulty during experiments for the convection heat transfer since the whole system requires heating equipment by the cooling one as well. The corrosion of the piping system and containers causes additional difficulties. From this point of view, CFD studies are therefore an important alternate assistant in the evaluation of the thermal-hydraulic characteristics of molten salts. To reveal the convective heat transfer characteristics of MS and check the validness of the well-known correlations pertaining to empirical convective heat transfer of them in the literature, some experimental [17-21] and numerical [1, 22-25] studies were carried recently. In these studies, such molten salts as $LiNO_3$, LiF-NaF-KF (FLiNaK), $NaNO_2$-$KNO_3$-$NaNO_3$, LiF-$BeF_2$-$ThF_4$-$UF_4$, $NaBF_4$-NaF, and LiF-$BeF_2$ (known as FLiBe) were considered.

The convective heat transfer with molten salt in different types of channel geometries; namely circular tubes, annular ducts, outside the tube bundles, and double pipes were conducted in these useful studies. The heat transfer enhancement in MSRs is important for reducing cost and fluid inventory. The amount of fuel out of the core in the MSRs should be adjusted to prevent a new critical mass. Due to the particular importance of the heat transfer enhancement and the requirement of a small volume to prevent the hold-up problem in MSRs power plants, compact heat exchanger arrangements are appropriate. A new concept which is called a Printed Circuit Heat Exchanger (PCHE), combines compactness, low-pressure drop, high effectiveness, and the ability to operate with a large pressure differential is manufactured by Heatric [26, 27].

A compact heat exchanger (CHE) design obtained by finned surfaces for an appropriate working fluid will gain a vital role in MSRs. Several types of molten salts have been investigated extensively by ORNL in support of the MSR Experiment and the MS Breeder Reactor Programs [26]. Due to its convenience for both MSRs fuel salt and coolant, FLiBe has been reported as a promising candidate during the fluid selection process. Therefore, the thermal and hydraulic characteristics of FLiBe flowing over the finned surface must be investigated. Regretfully, there isn't any work that reports neither experimental nor numerical findings for a louvered fin flat-tube compact heat exchanger associated with MSRs in open sources. Computational Fluid Dynamics (CFD) can be used to fulfill the lack of data and information in the field.

In this study, a heat exchanger has been considered as a louvered fin flat-tube compact type to get high heat transfer rates with higher heat transfer areas by considering the whole explanations up here. FLiBe (LiF-BeF$_2$), a frequently selected molten salt in MSRs and MSFRs, has been selected as the working fluid. The benchmark study investigates the steady-state thermo-hydraulic behavior of FLiBe flowing through a compact heat exchanger with multi louvered fins in the cooling loop of an MSR for a low-velocity range. The heat exchanger could be a primary or a secondary heat exchanger. The numerical tests are conducted for low Reynolds numbers of 100 – 500 to simulate unusual conditions faced with such as coolant pump failure. A series of numerical tests are conducted for the louver pitch with different fin pitches such as 1.5, 2.0, and 2.5 mm and fin louver angles as 20º, 24º, 28º, 32º, 36º. The velocity profiles, temperature, and pressure fields are obtained and to get a performance comparison for FLiBe as the coolant in an MSR, the heat transfer coefficient and pressure drop data are reported in terms of Colburn *j*-factor and Fanning friction factor *f*, and the volume goodness factor (*j/f*$^{1/3}$). Though there is a growing concern about the usage of molten salts in nuclear reactors, there are very few investigations on the alternate heat exchangers that can be used in these systems. This particular investigation is believed that it will contribute to the MSR literature. In this paper, the geometry of the louvered finned flow channel is introduced first, a CFD model to simulate the convection of FLiBe over the louvered fins was explained, and then validated results are presented.

## 2. Material and Method

As mentioned previously, a compact heat exchanger with the louvered fin on a flat tube and FLiBe as the working fluid is examined numerically in this particular work to provide preliminary outcomes regarding its applicability in the field of MSRs. Though numerous working fluids can be used in MSRs, the FLiBe is chosen in the study. Considerations for material selection are explained in the first part of this section. The detail of the mathematical method is given in the second part of this section.

### 2.1 Considerations for coolant selection

To improve the efficiency of the reactors, several types of working fluids have been proposed and used by researchers. Since a variety of molten salts can be used in the reactors, the selection of the right working fluid is crucial. To ease the selection, a series of the figure of merits (FOM) has been suggested by Kim et al. [28] as regards to the general characteristics that are required in the coolant (Table 1). By considering the suggested FOMs for each general characteristic of the coolants, molten salts considered in MSRs could be compared from a better aspect. The physical properties of typical molten salts used in the reactors and FOM's are summarized in Table 2. The original table could be addressed in ref. [28] in which results of not only the molten salts but also gas and liquid metals could be found as well. In Table 2, in addition to the typical molten salts reported by Kim et al. [28], the physical properties and FOM's of the FLiBe is also enlisted at the bottom in the present study. The first one, $FOM_{ht}$, among the suggested FOMs is regarding the heat transfer performance of the coolant. As for the evaluation of the findings of suggested FOM's, Kim et al. [28] reported that $FOM_{ht}$ would get higher values while the remaining ones ($FOM_p$, $FOM_{cv}$, $FOM_{ccv}$ and $FOM_{hl}$) would get lower values to achieve a better performance. From the highlighted point of view, it is seen that, FLiBe is superior than the alternative molten salts. Though it has moderate $FOM_{ht}$, it has the lowest FOM's in remaining four FOMs.

**Table 1.** Definition of the figure of merits (FOMs)

| Definition | FOM | |
|---|---|---|
| Heat transfer performance of coolant where, $FOM_{ht,0}=(k)^{0.6}(\rho)^{0.58}(C_p)^{0.4}(\mu)^{-0.47}$ | $FOM_{ht} = \dfrac{(k)^{0.6}(\rho)^{0.58}(C_p)^{0.4}(\mu)^{-0.47}}{FOM_{ht,0}}$ | (1) |
| Pumping power of coolant where $FOM_{p,0}=(\rho)^{-2}(C_p)^{-2.8}(\mu)^{02}$ | $FOM_p = \dfrac{(\rho)^{-2}(C_p)^{-2.8}(\mu)^{02}}{FOM_{p,0}}$ | (2) |
| Volume of coolant where $FOM_{cv,0}=(\rho)^{-0.84}(C_p)^{-1.16}(\mu)^{0.1}$ | $FOM_{cv} = \dfrac{(\rho)^{-0.84}(C_p)^{-1.16}(\mu)^{0.1}}{FOM_{cv,0}}$ | (3) |
| Volume of structural materials where $FOM_{ccv,0}=(P)(\rho)^{-0.84}(C_p)^{-1.16}(\mu)^{0.1}$ | $FOM_{ccv} = \dfrac{(P)(\rho)^{-0.84}(C_p)^{-1.16}(\mu)^{0.1}}{FOM_{ccv,0}}$ | (4) |
| Heat loss of the coolant where $FOM_{hl,0}=(k)^{0.6}(\rho)^{0.34}(C_p)^{0.06}(\mu)^{-0.44}$ | $FOM_{hl} = \dfrac{(k)^{0.6}(\rho)^{0.34}(C_p)^{0.06}(\mu)^{-0.44}}{FOM_{hl,0}}$ | (5) |

**Table 2.** Comparison of FOMs for different coolants

| Coolant | Melting Point (°C) | $k$ (W/mK) | $\rho$ (kg/m$^3$) | $C_p$ (J/kgK) | $\mu$ (Pa·s) | $P$ (atm) | $FOM_{th}$ - | $FOM_p$ - | $FOM_{cv}$ - | $FOM_{ccv}$ - | $FOM_{hl}$ - |
|---|---|---|---|---|---|---|---|---|---|---|---|
| Water (25°C) | 0 | 0.61 | 997.05 | 4181 | 0.00089 | 1 | 1.00 | 1.00 | 1.00 | 1.00 | 1.00 |
| LiF-NaF-KF | 454 | 0.92 | 2020 | 1886 | 0.0029 | 1 | 0.80 | 2.87 | 1.57 | 1.57 | 0.92 |
| NaF-ZrF4 | 500 | 0.49 | 3140 | 1173 | 0.0051 | 1 | 0.45 | 5.02 | 1.98 | 1.98 | 0.56 |
| KF-ZrF4 | 390 | 0.45 | 2800 | 1046 | 0.0051 | 1 | 0.38 | 8.70 | 2.50 | 2.50 | 0.51 |
| LiF-NaF-ZrF4 | 436 | 0.53 | 2920 | 1233 | 0.0069 | 1 | 0.40 | 5.36 | 2.05 | 2.05 | 0.50 |
| LiCl-KCl | 355 | 0.42 | 1520 | 1198 | 0.00115 | 1 | 0.55 | 14.99 | 3.07 | 3.07 | 0.76 |
| LiCl-RbCl | 313 | 0.36 | 1880 | 890 | 0.0013 | 1 | 0.47 | 23.08 | 3.67 | 3.67 | 0.70 |
| NaCl-MgCl2 | 445 | 0.50 | 1680 | 1096 | 0.00136 | 1 | 0.58 | 16.28 | 3.18 | 3.18 | 0.81 |
| KCl-MgCl2 | 426 | 0.40 | 1660 | 1160 | 0.0014 | 1 | 0.50 | 14.31 | 3.02 | 3.02 | 0.70 |
| NaF-NaBF4 | 385 | 0.40 | 1750 | 1507 | 0.0009 | 1 | 0.71 | 5.66 | 2.04 | 2.04 | 0.88 |
| KF-KBF4 | 460 | 0.38 | 1700 | 1305 | 0.0009 | 1 | 0.64 | 8.98 | 2.47 | 2.47 | 0.84 |
| RbF-RbF4 | 442 | 0.28 | 2210 | 909 | 0.0009 | 1 | 0.54 | 14.63 | 3.01 | 3.01 | 0.75 |
| FLiBe LiF – BeF2 | 459 | 1.00 | 1940 | 2414.17 | 0.0056 | 1 | 0.67 | 1.78 | 1.30 | 1.30 | 0.73 |

## 2.2 Mathematical Method

Technical drawing of considered louvered fin geometry is shown in Fig. 1. The fin material is Hastelloy-N which is a nickel-base alloy developed for MSRs by ORNL and the International Nickel Company. The performance of the louvered fin is analyzed for the fin pitches of $F_p$=1.5, 2.0, and 2.5 mm and louver angles of $L_\alpha$=20º, 24º, 28º, 32º, 36º while the louver pitch and fin thickness are constant as $L_p$=1.7 mm and $t$=0.1 mm, respectively.

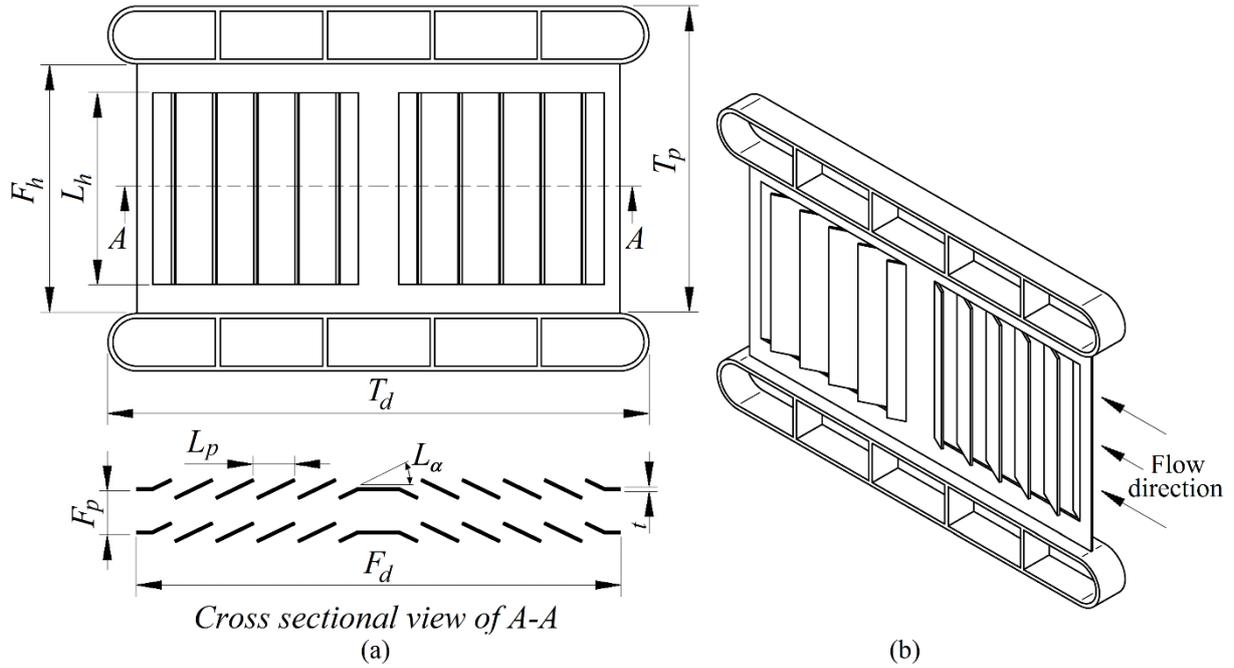

**Fig. 1.** Technical drawing of the considered louvered fin
(a) 2-D cross-sectional view of *A-A* (b) 3-D appearance

The Reynolds number for the coolant side based on the louvered pitch is given as

$$Re_{Lp} = \frac{u_c L_p}{v} \quad (6)$$

where

$$u_c = u_{in} \frac{A_{fr}}{A_c} \quad (7)$$

The heat transfer coefficient of the coolant side is determined by using Eq. 8.

$$h = \frac{\dot{Q}}{A_a \Delta T_m} \quad (8)$$

where the heat transfer rate and logarithmic mean temperature difference are defined as

$$\dot{Q} = \dot{m}c_p(T_o - T_i) \tag{9}$$

$$\Delta T_m = \frac{T_i - T_o}{\ln[(T_s - T_o)/(T_s - T_i)]} \tag{10}$$

In heat exchanger designs, the overall performance criteria are used instead of direct use of heat transfer coefficient. The pressure gradient is one of the key design parameters to be considered when the molten salt has flowed over the louvered fins. The thermal and hydraulic performance of the louvered fin heat exchanger is presented in terms of the Colburn factor ($j$) and the friction factor ($f$), which are given by Eqs. 11-12.

$$j = StPr^{2/3} \tag{11}$$

$$f = \frac{2\Delta P}{\rho u_c^2} \cdot \left(\frac{A_c}{A_a}\right) \tag{12}$$

where

$$St = \frac{h}{\rho u_c c_p} \tag{13}$$

The area goodness factor of $j/f$ and the volume goodness factor of $j/f^{1/3}$ are used for expressing new sights about the overall performance of the louvered fin heat exchanger.

## 2.3 Numerical Method

By regarding the fact that the maximal velocity in each section of the heat exchanger needs to be limited to 5 m/s to avoid erosion and corrosion [13]. The FLiBe flow was assumed to be laminar for $Re_{Lp} \leq 1300$ [29]. The mass, momentum, and energy equations for 2-D, steady, laminar, and incompressible flow are given respectively in Eqs. 14-16.

$$\nabla \cdot V = 0 \tag{14}$$

$$(V \cdot \nabla)V = -\frac{1}{\rho}\nabla P + \upsilon \nabla^2 V \tag{15}$$

$$(V \cdot \nabla)T = \alpha \nabla^2 T \tag{16}$$

In modeling and analysis of the primary and secondary heat exchangers with louvered fins in MSRs, the expected properties of FLiBe and Hastelloy-N were assumed to be constant as shown in Table 3.

**Table 3.** Properties of FLiBe and Hastelloy-N

| Material | $\rho$ (kg/m³) | $c_p$ (J/kgK) | $k$ (W/mK) | $\mu$ (kg/ms) | $Pr$ |
|---|---|---|---|---|---|
| Hastelloy-N [13] | 8860 | 578 | 23.6 | - | - |
| FLiBe [15] | 1940 | 2414.17 | 1.0 | 0.0056 | 13.525 |

## 2.4 Boundary Conditions

The considered 2-D computational domain is shown in Fig. 2. The interface of the louvered fins is defined as periodic surfaces due to the periodicity of the louvered fins and a no-slip boundary condition is applied on the fin surfaces. The constant temperature of 973 K is assumed on the fin wall since the melting temperature of Hastelloy-N is around 1300-1400°C. The coolant temperature at the inlet of the louvered fins is 773 K since the fuel and coolant temperatures are determined in the range of 490-800°C [30]. The inlet velocity of FLiBe is 0.15-0.80 m/s to provide laminar flow conditions at the inlet of the louvered fins and the gauge pressure of 0 Pa is applied to the exit of the louvered fins. The mass, momentum, and energy equations of the numerical model are solved by using the second-order upwind scheme with Ansys Fluent 14 [31]. The simple algorithm is used to get the shortest running for the coupling of velocity and pressure fields of the coolant over the louvered fins [30].

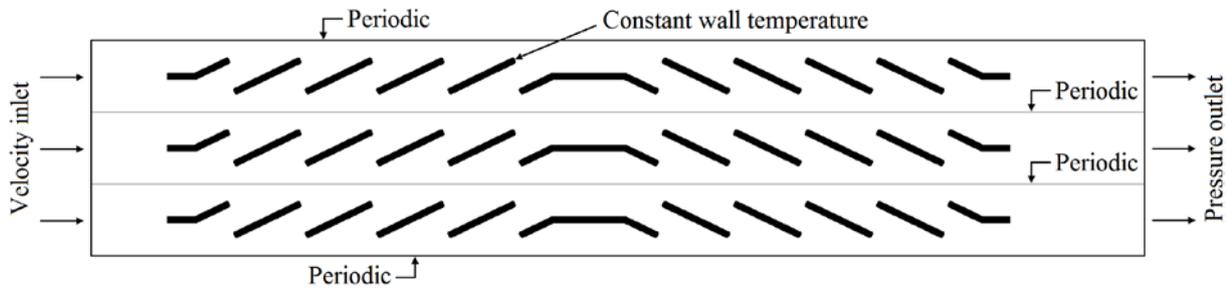

**Fig. 2.** The 2-D numerical model and the boundary conditions

## 2.5 Validation of the Computational Domain

The four different sizes of quadrilateral cells were used to get the mesh independence of the 2-D numerical model as shown in Fig. 3a. It was found that the difference in the solutions were about 0.16% and 0.96% respectively for $j$ and $f$ for meshes with more than 90741 nodes. Due to the severe lack of information about the case in which FLiBe molten salt flows over compact heat exchangers with louvered fins, it is very challenging to validate the computational results of the study with literature findings. The major difficulty is not only the molten salt-compact heat exchanger couple but also the very narrow flow depth of the flat tube used to manufacture the CHEX. Regarding these challenges, a basic comparison is used for the validation of the model used in the study. Since any experimental results on this particular topic do not exist, the reliability of the model is tested by a numerical study reported by Perrotin and Clodic [32] for Fig. 3b. Though several conditions have been investigated in the referred study, the model with $F_p$=1.30 mm, $L_p$=0.95 mm, $L_\alpha$=20°, $F_d$=16.5 mm is the most suitable and comparable one with our present model. As it could be seen there is a good agreement between the CFD results of the model and numerical results of Perrotin and Clodic [32].

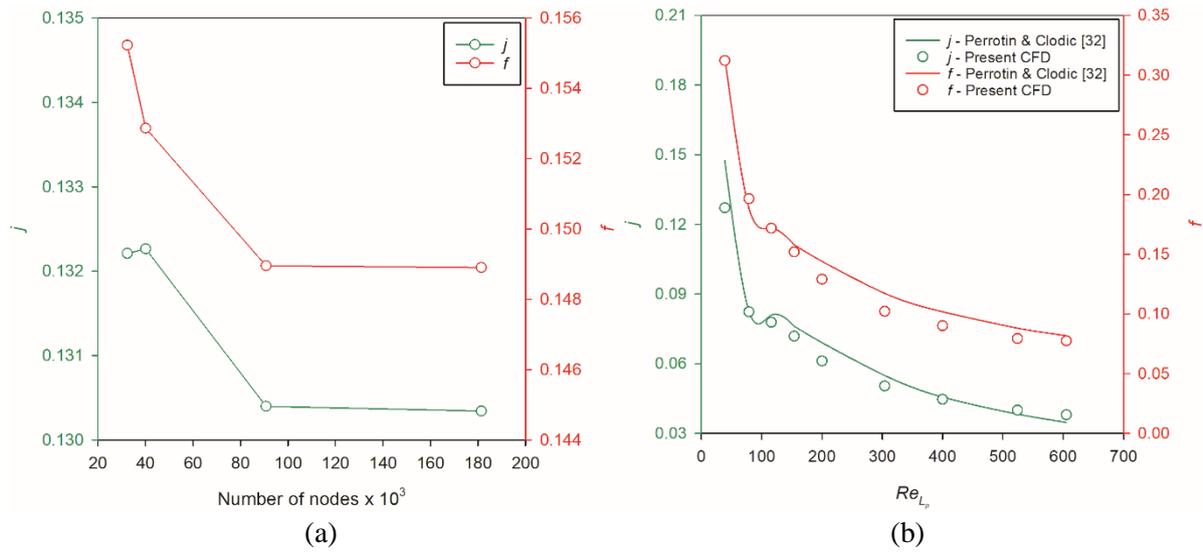
**Fig. 3.** (a) Mesh independency (b) Comparison of the results with the literature

## 3. Results and Discussion

In this work, the thermo-hydraulic performance of the molten salt FliBe is investigated numerically. The investigation is conducted for varying fluid velocities in the range of 0.1 m/s - 0.8 m/s, fin pitches in the range of 1.5 - 2.5 mm, and fin louver angles in the range of 20° - 36°. The findings of the simulations are presented either in dimensional form (heat transfer coefficient vs. frontal velocity, pressure drop vs. frontal velocity) or non-dimensional form (Colburn factor vs. Re number, friction factor vs. Re number). As for the initial step, the contour plots for streamline, pressure, and temperature are demonstrated to provide a better perspective on the flow characteristics of the studied molten salt. Then the thermo-hydraulic performance of the FLiBe is presented by the diagrams that are driven for heat transfer coefficient, pressure drop, Colburn *j*-factor, and friction factor.

The streamlines which are tangent to the local velocity vectors during the flow of the FLiBe are presented in Fig. 4 for the fin pitch $F_p$=1.5 mm and for the angles $L_\alpha$=20° - 32° where $Re_{Lp}$=100. These lines are used to observe the inclination of the flow direction during the analyses, as regards to the given figures, when the stream becomes stronger in any direction, the color turns to red from blue. While blue demonstrates a weaker stream, red corresponds to a stronger flow stream. As it could be seen from the schematic, when the louver angle is small ($L_\alpha$=20°), the molten salt tends to flow through the space between louver fins which causes a lesser interaction between the FLiBe and fin surfaces. On the contrary, the interaction between fin surfaces and FLiBe improves as the louver angle becomes steep. As it is presented in Fig. 4, the streamlines of molten salt flow become stronger, and the color of the streams turns red when FLiBe flows over the fins. It can be noted that, as the angle is steeper the flow of the molten salt can be more directed to louver fins and a "louver directed flow" can be obtained vice versa if the angle is moderate, molten salt's flow turns to a duct directed flow instead.

The pressure changes along with the louver fins and consequently the drop between the upstream and downstream can be observed from Fig. 5. As it could be seen, as the louver angle turns from $L_\alpha$=20° to 32°, the pressure changes from upstream to downstream increases. As regards to the presented figure, the pressure of the FLiBe is high at the upstream and rises locally in particular at the leading edge in both presented louver angles. The rise at the leading edge is bigger when the fin is steep since the resistance against the flow is higher. As the molten salt flows over the fin, the pressure falls along the flow path and becomes the lowest at the downstream. The pressure difference between the inlet and exit of the molten salt flow is higher when the fin is steeper, and this variance descends as the louver angle becomes more moderate where the resistance against the flow is lower.

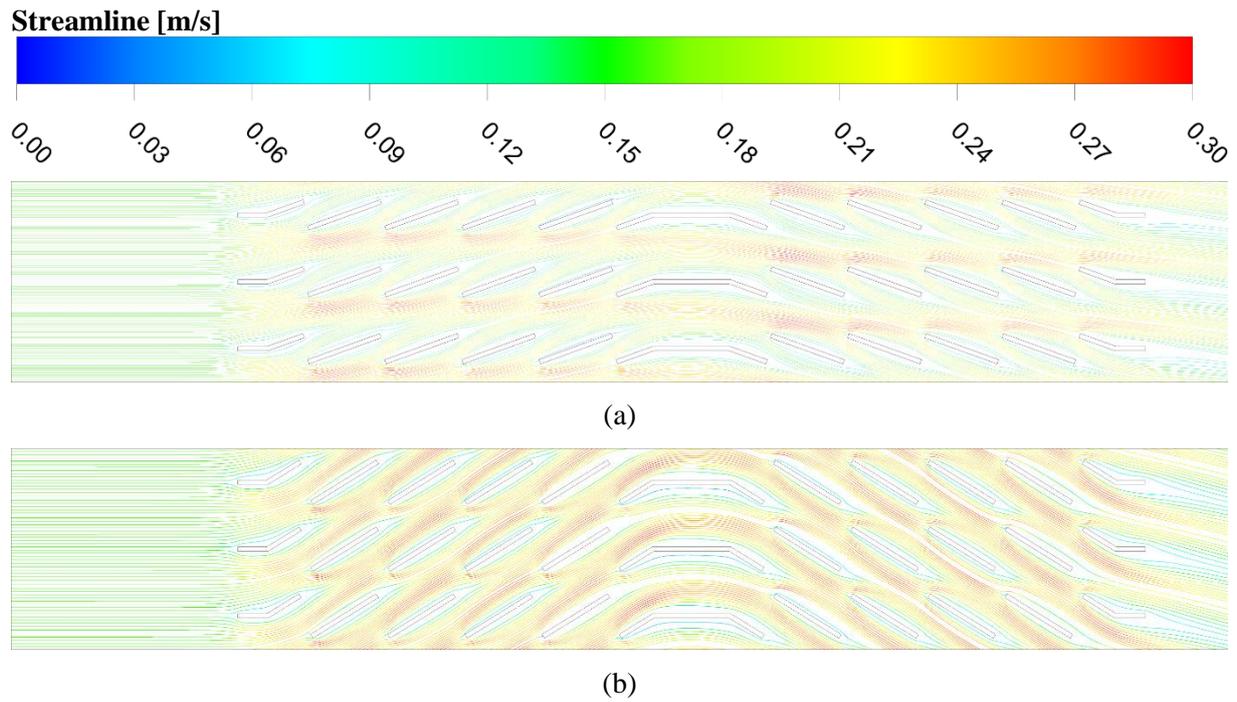

**Fig. 4.** Streamlines of FLiBe flowing through the louvered fins for $F_p$=1.5 mm and $Re_{Lp}$=100
(a) $L_\alpha$=20° (b) $L_\alpha$=32°

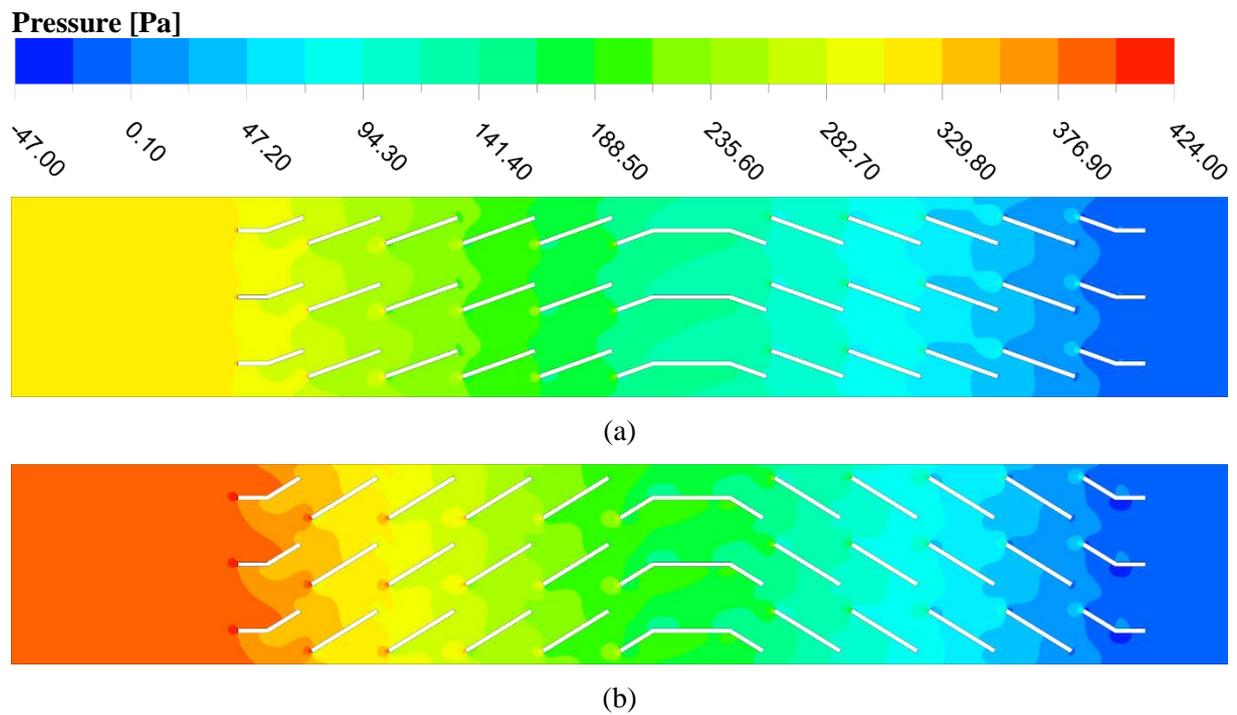

**Fig. 5.** Pressure fields of FLiBe flow through the louvered fins for $F_p$=1.5 mm and $Re_{Lp}$=100
(a) $L_\alpha$=20° (b) $L_\alpha$=32°

The temperature distribution along the louver fins is presented in Fig. 6 for the $F_p$=1.5 mm and $L_\alpha$=20°-32° at $Re_{Lp}$=100. Temperature rises of the molten salt from the upstream to the downstream can be seen clearly in the presented contour plots. While the FLiBe's temperature at the upstream is 773 K, at the downstream temperature can rise to 853-873 K locally as regards to heat transfer over the fins to the molten salt. Due to the interaction between the fin surface and FLiBe, more heat is transmitted to the molten salt in the neighboring regions of the fins in which temperature can go up to 950 K as it is plotted in the figures by red color. In accordance with the angle increment, the molten salt gains more heat and the temperature of the molten salt ascends gradually as the $L_\alpha$ rises from 20° to 32°. The temperature of the FLiBe is higher when the louver angle is steep, the major reason for this increment relies on the flow characteristics which are presented in Fig.4. When the louver angle rises, the flow turns from duct directed to louvered directed which improves the interaction between the fins and molten salt, and consequently, the heat transfer is improved.

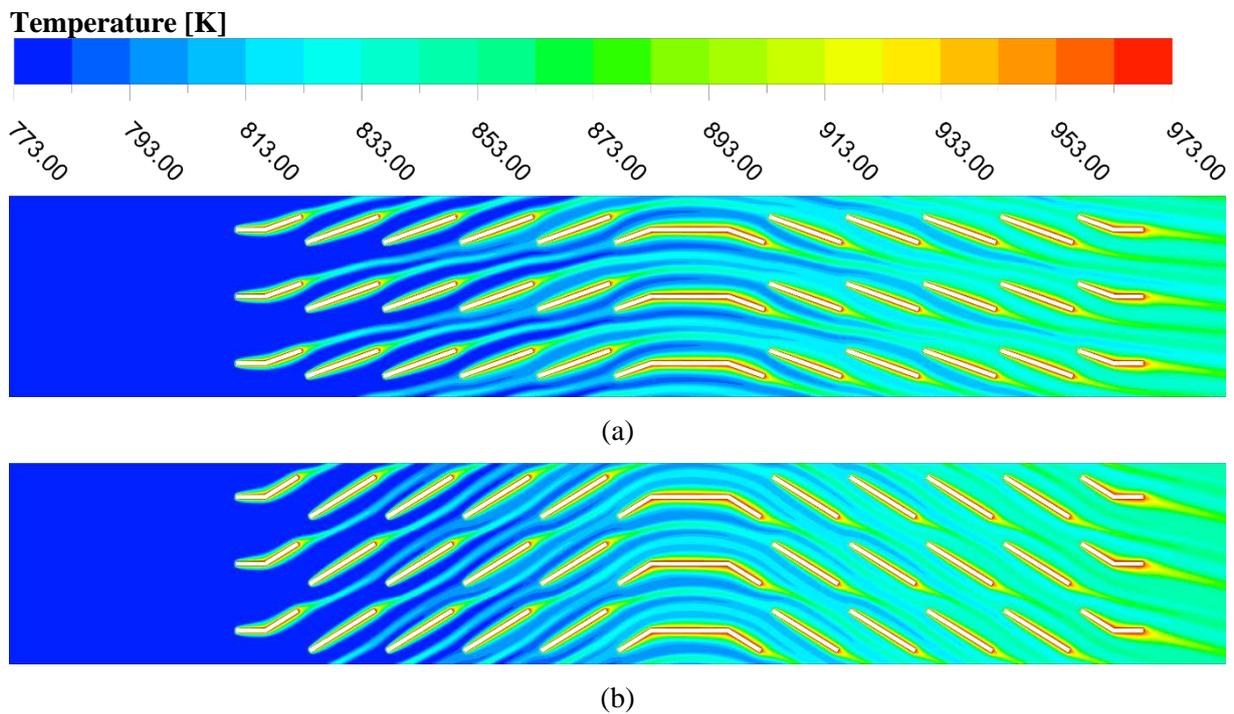

**Fig. 6.** Temperature fields of FLiBe flow through the louvered fins for $F_p$=1.5 mm and $Re_{Lp}$=100
(a) $L_\alpha$=20° (b) $L_\alpha$=32°

The evolution of the same parameters (streamlines, pressure, and temperature) by the contour plots are presented for the spacing $F_p$=2.5 mm in Fig 7-9. To present the findings from a different aspect. A wider fin spacing ($F_p$=2.5 mm) and higher Reynolds number ($Re_{Lp}$=400) are chosen in the mentioned figures and temperature, pressure, and streamline contours are plotted for these particular cases. Once again, the louver angles $L_\alpha$=20°-32° are chosen to demonstrate the impact of the angle change on the flow characteristics of the molten salt. Similar trends are observed in each figure for the corresponding parameters, such as, while the 'duct directed' flow is observed for $L_\alpha$=20°, it turns to 'louvered directed flow' when the louver angle becomes steep ($L_\alpha$=32°) in Fig 7.

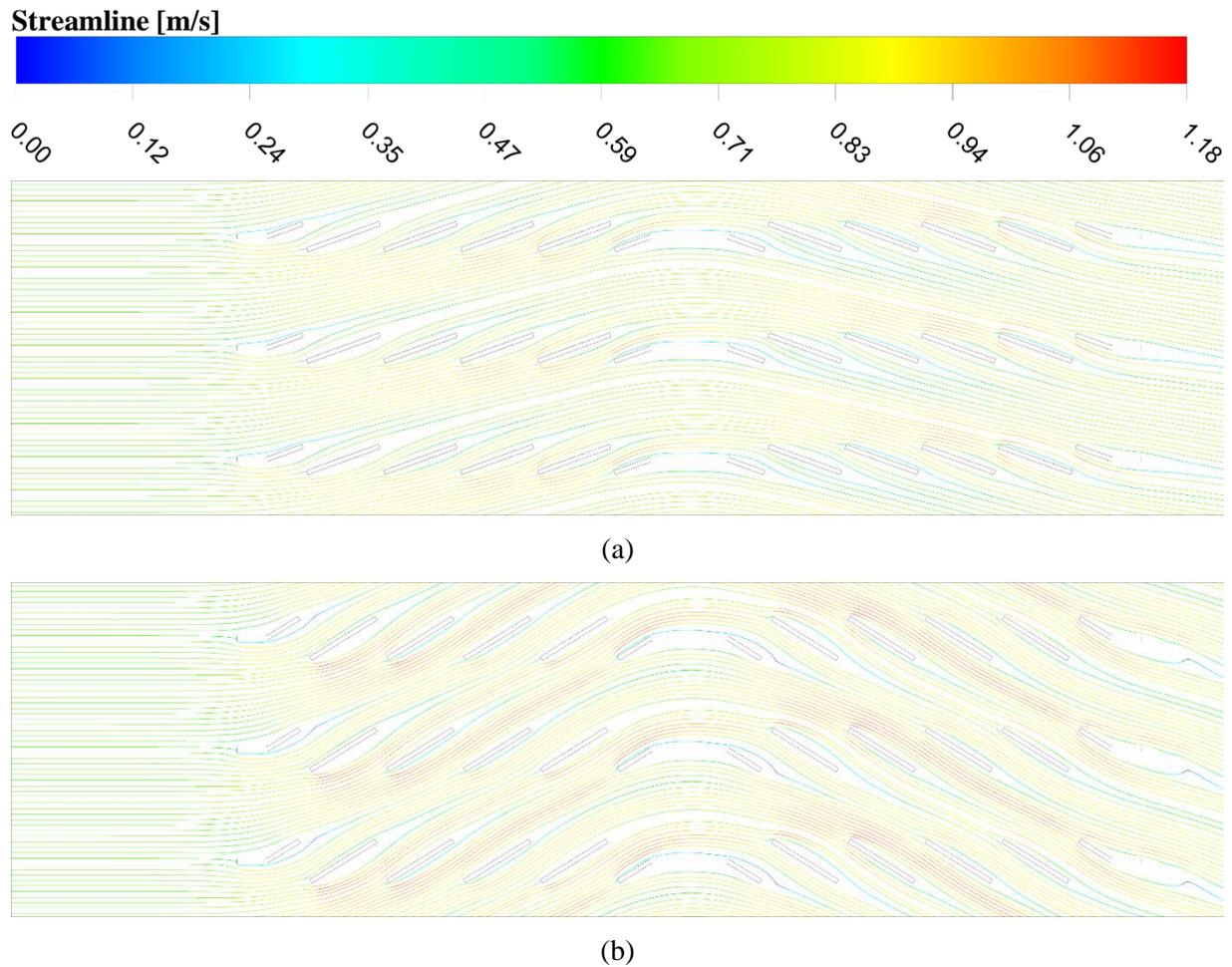

(a)

(b)

**Fig.7.** Streamlines of FLiBe flowing through the louvered fins for $F_p$=2.5 mm and $Re_{Lp}$=400
(a) $L_\alpha$=20° (b) $L_\alpha$=32°

The pressure reduces across the flow direction, the higher pressure is observed at the upstream and lower pressure is seen at the downstream in both angles $L_\alpha$=20°-32°. Since the resistance is more when the louver fins are steep, the pressure at the upstream is higher $L_\alpha$=32° and therefore the pressure difference between inlet and exit is more as it could be seen in Fig. 8. The temperature distribution over the molten salt increases along the flow direction as well in Fig. 9 during the interaction of the molten salt with fins. Though identical trends are observed, some notable features are detected as well in the demonstrated figures. For instance, even though flow directions are identical, streamlines are sparse due to the higher fin pitch used in this model (Fig. 7). The major reason for this remarkable change is the space between louver fins indicated by the pitch which is 2.5 mm instead. As the space grows, the bigger portion of the molten salt flows through that space instead of flowing over the fins and flow becomes more "duct directed" instead of "louvered directed". This is also the major reason for the worsening thermal performance observed.

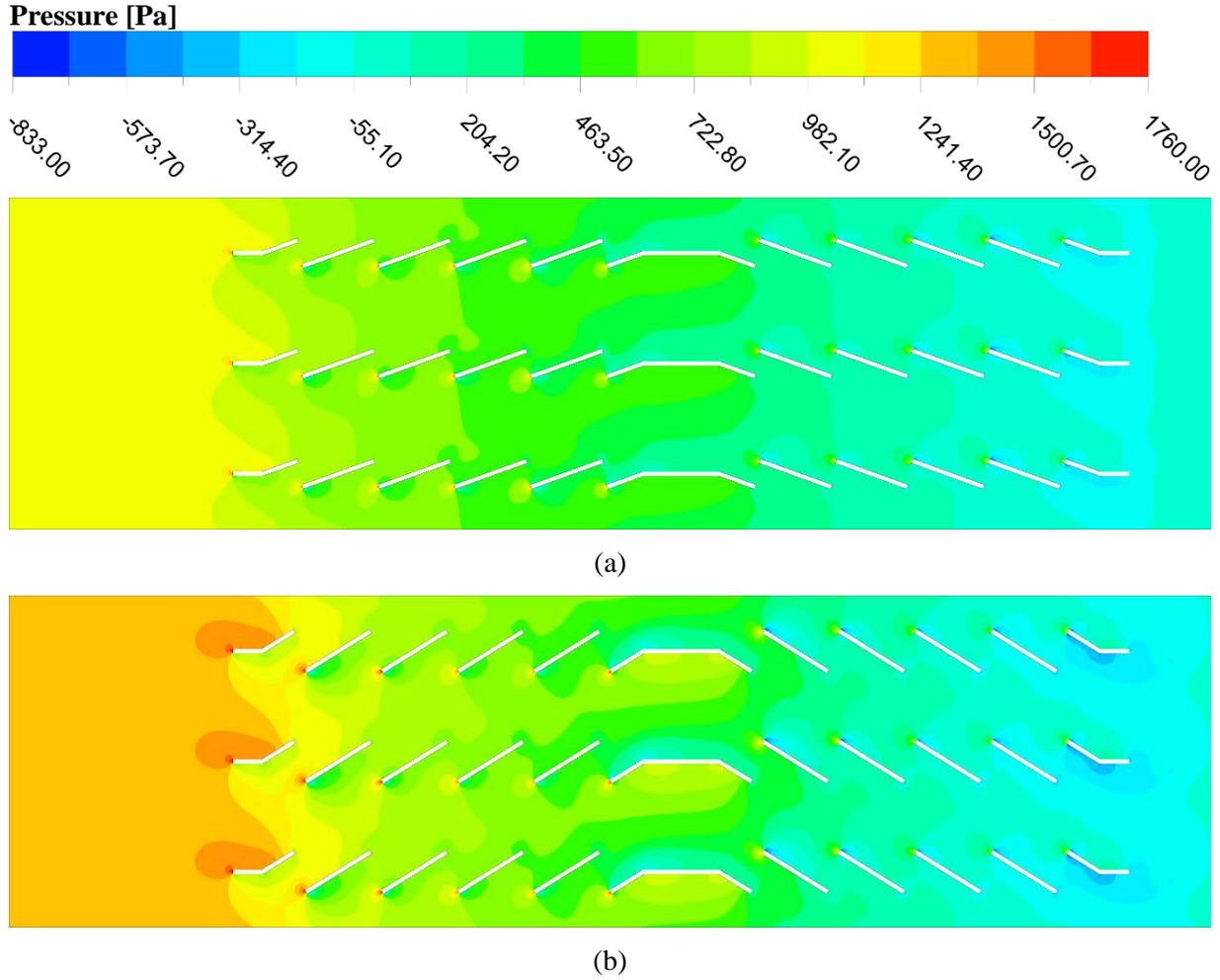

**Fig. 8.** Pressure fields of FLiBe flowing through the louvered fins for $F_p$=2.5 mm and $Re_{Lp}$=400
(a) $L_\alpha$=20° (b) $L_\alpha$=32°

Fig. 9. As it is seen, the region that gains heat from the fins is very limited since the interaction between the FLiBe and fins gets worse. As the contact reduces and molten salt has more free space to flow, the heat transfer from the fins to the FLiBe gradually descends and the temperature plot has more blue regions which refer to inlet temperature.

Heat transfer coefficient (*h*) and pressure drop (*ΔP*), variation as regards to molten salt frontal velocity is presented in Fig.10 for three different fin pitches ($F_p$=1.5, 2.0, and 2.5 mm). As the velocity increases, the heat transfer performance improves regardless of angle change and fin pitch variation of the louver fins. The left-hand side of the figures pertains to heat transfer coefficients, which refers to the curve cluster at the top, while the right-hand side pertains to pressure drop, which refers to the curve cluster at the bottom of the figures. Among the investigated louver angles, the $L_\alpha$=36° provides the highest heat transfer which gets higher values as the frontal velocity ascends. The major reason for this improvement relies on the flow direction change provided by the variation of the louver angle as presented before with the contour plots. Since the molten salt flow becomes more louver-directed when the angle is steep, it enhances the interaction between the FLiBe and fins which leads to a thermal performance increment. The maximum h is obtained as 27000 W/m²K at this angle ($L_α=36°$) and for the velocity of 0.8 m/s among all tested cases, while the lowest is obtained as 12000 W/m²K at ($L_α=20°$) and for the velocity of 0.1 m/s. Even though, there is a noticeable difference between the angle $L_α=36°$ and the remaining angles, this difference is not obvious as the $L_α$ decreases, so h values begin to converge at lower angle values and starts to present identical evolution trends by the frontal velocity.

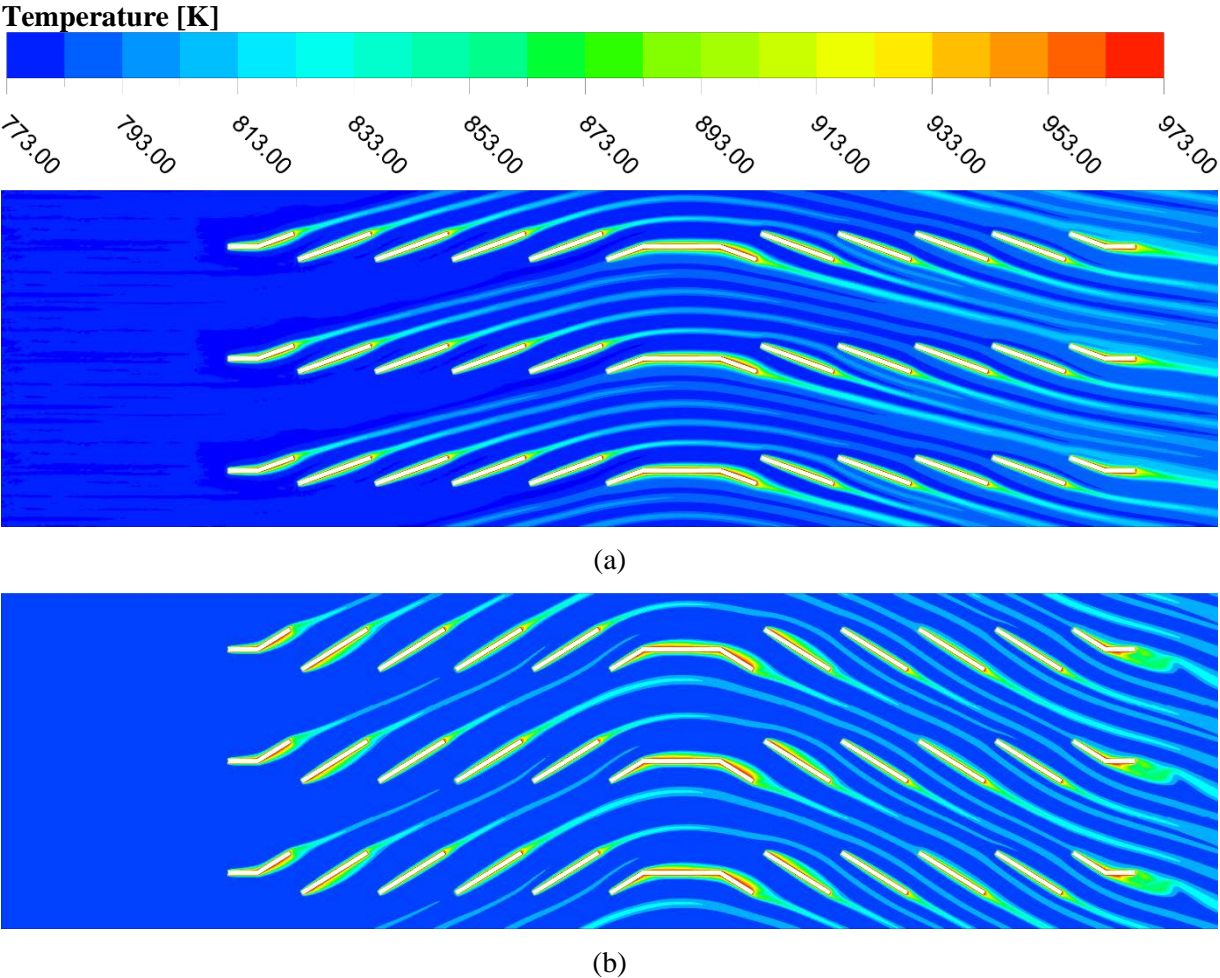

**Fig. 9.** Temperature fields of FLiBe flowing through the louvered fins for $F_p$=2.5 mm and $Re_{Lp}$=400 (a) $L_α=20°$ (b) $L_α=32°$

The evolution characteristics are more obvious for pressure drop (Δ$P$) curves in the same figure (Fig. 10a). As the louver angle grows, the fins become steep and as a result of increasing pressure at the leading edge, the difference between the upstream and downstream become more obvious. The highest pressure drop is obtained for the highest angle at the maximum frontal velocity while the lowest is observed for the lowest angle and the lowest velocity.

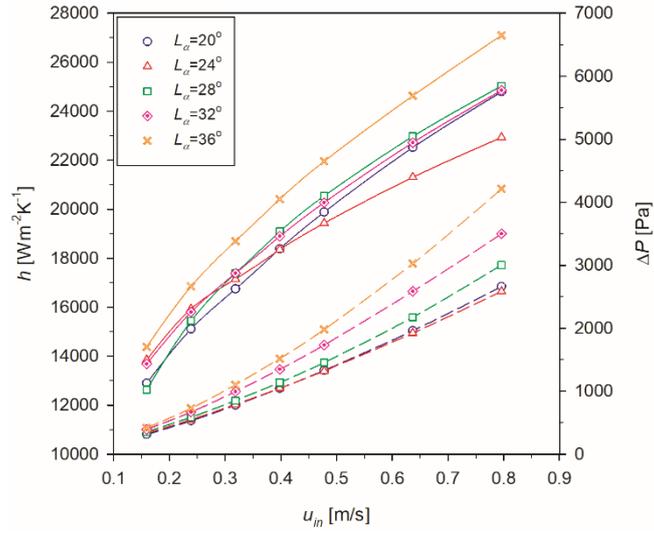

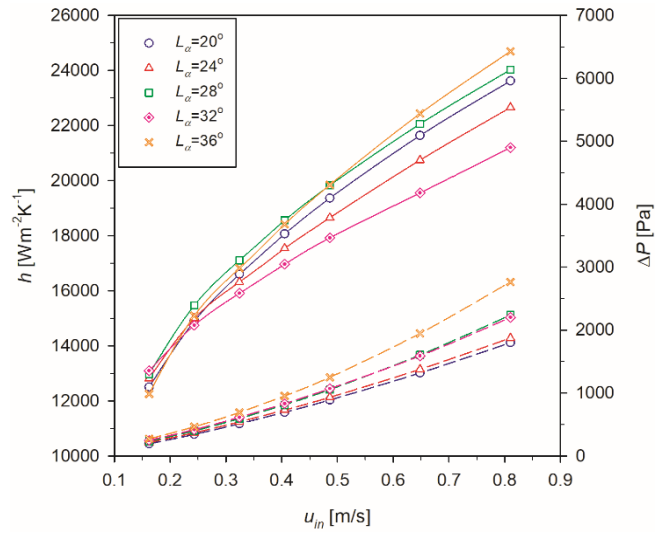

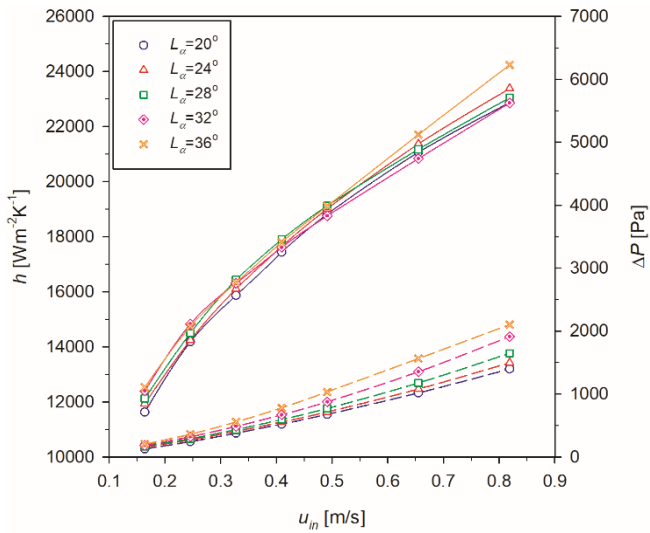

**Fig. 10.** Deviation of heat transfer coefficient (*h*) and pressure drop (*ΔP*) with respect to frontal velocity for FLiBe ($u_{in}$) (a) $F_p$=1.5 mm (b) $F_p$=2.0 mm (c) $F_p$=2.5 mm

The possible reason for the reported change can be related to the angle. As the angle increases, the resistance against molten salt flow ascends too, which leads to a higher pressure drop. Similarly, as the velocity increases, the pressure drop rises as well. Contradiction to this, as the frontal velocity reduces, pressure drop reduces and when this fall approaches about 0.1 m/s, almost all angles show a similar pressure drop across the flow over the fins at this particular velocity. It is seen that; the resistance that emerges by the angle rise doesn't have an impact at lower frontal velocities as in the higher ones. Similar evolution trends for heat transfer ($h$) and pressure drop ($\Delta P$) can be observed for $F_p$=2.0 mm in Fig. 10b. As it is seen, either $h$ or $\Delta P$ increases as the frontal velocity of the molten salt ascends, in addition, both of them rise as the louver angle steeps. Unlike Fig. 10a, h curves of varying louver angles are more obvious at higher frontal velocities. The gradual increment of $h$ plots in each louver angle is more clear than the previous fin pitch. The highest $h$ is obtained around 25000 W/m²K when the $L_\alpha$= 36° and $v$=0.8 m/s. As for $\Delta P$, the difference between the plots becomes smaller at this fin pitch but is still apparent. The angle change from $L_\alpha$=32° to 28° doesn't have much impact on pressure drop and is located between $L_\alpha$=36° and 24°. Even the $L_\alpha$=20° has the least $\Delta P$ among the analyzed angles, $\Delta P$ at $L_\alpha$=24° is very close and there is a vague difference between these two. Fig. 10c demonstrates the heat transfer coefficient ($h$) and pressure drop ($\Delta P$) evolution for the fin pitch $F_p$ = 2.5 mm. Though the trend is identical with the earlier findings, h's are becoming closer than earlier plots which are belonging to fin pitches of 1.5 mm and 2.0 mm. Since the space between the louver fins increases, the duct-directed flow becomes more prominent for the molten salt and as a result, the interaction with the fins descends and thermal performance reduces. Associated with the previous reason, the impact of the angle on the thermal performance descends as well and $h$ could be able to reach to 24000 W/m²K, though the $\Delta P$ trend and the order among the angles are similar, the values of $\Delta P$ and curves are plotted on a narrower range, unlike the earlier figures. While pressure drops can go beyond 3000 Pa and be able to reach 4000 Pa for the smaller fin pitches, when the $F_p$ = 2.5 mm, the $\Delta P$ barley exceeds 2000 Pa.

The Colburn $j$ factor and friction factor $f$ evolution as regards to Re number for three different fin pitches ($F_p$=1.5, 2.0, and 2.5 mm) are given in Fig. 11 to see the impact of the angle and velocity on the performance from a non-dimensional perspective. The Colburn $j$ factor and friction factor $f$ reduce as the Re number increases as regards to the nature of these factors' definitions. The highest Colburn $j$ factor is observed for the biggest louver angle in accordance with the earlier findings. The Colburn $j$ factor plots begin to cross over when the louver angle is less than $L_\alpha$=36° and it seems below this value, angle degree doesn't have much impact on thermo-hydraulic performance when fin pitch is 1.5mm (Fig 11a). Even though there is a vague difference between the curve for $L_\alpha$=24° and the rest of them, the variance cannot be reported as a remarkable one. The lowest performance for the Colburn $j$-factor can be marked at this particular condition at around 0.035.

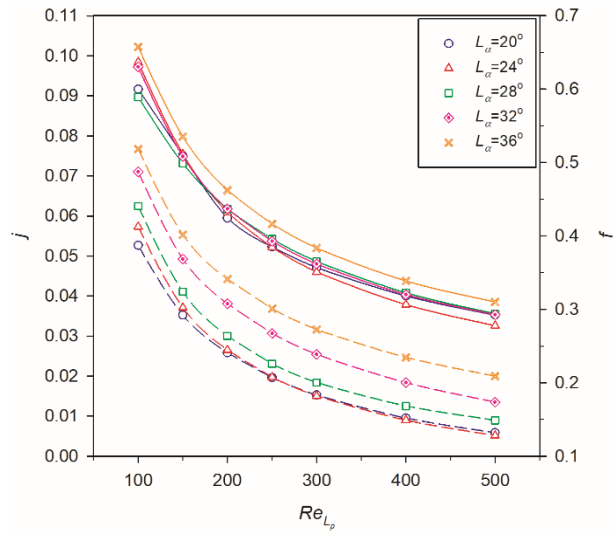

(a)

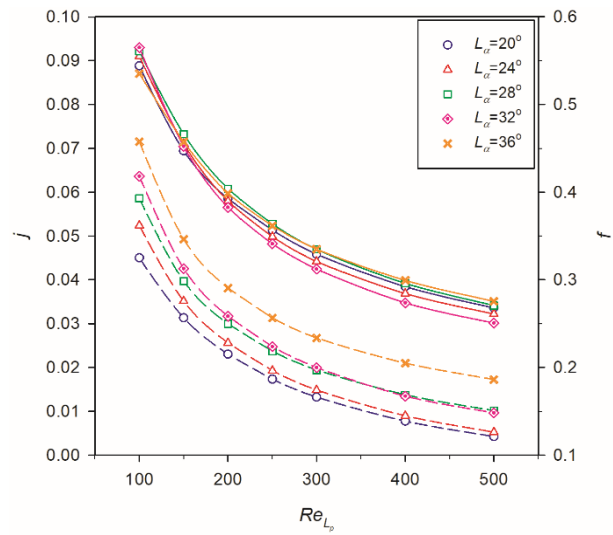

(b)

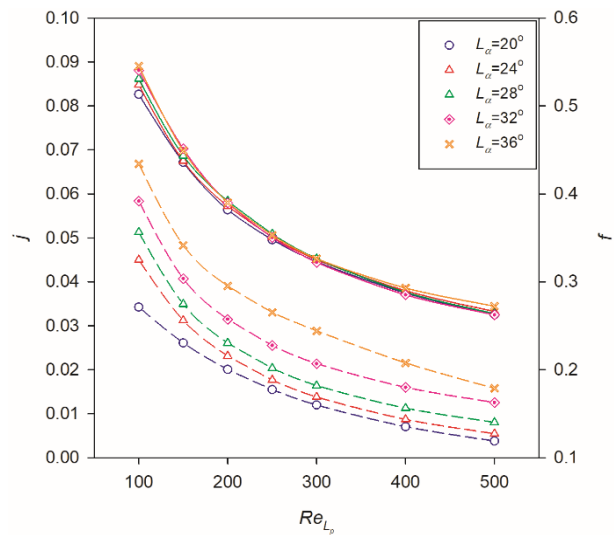

(c)

**Fig. 11.** Deviation of Colburn $j$- and $f$ factors with respect to frontal velocity for FLiBe (a) $F_p$=1.5 mm (b) $F_p$=2.0 mm (c) $F_p$=2.5 mm

The flow direction which is oriented by the louver angle change has an unavoidable impact on the thermal performance. As the angle steeps the flow turns to a louver-directed one and improves the interaction of the fin and molten salt and consequently improves the performance. As for the friction factor $f$, the evolution curves are more observable than the Colburn $j$-factor curves. The overall trend of the friction factor $f$ reduces by increasing $Re_{Lp}$ number which can be observed in all investigated louver angles. The highest $f$ factor is recorded as 0.52 for the $L_\alpha=36°$ when $Re_{Lp}$ is 100 while the lowest is obtained for $L_\alpha=20°$-24° when $Re_{Lp}$ is 500. As it is seen, the friction factor at the louver angles $L_\alpha=20°$-24°, becomes very similar when $Re_{Lp}$ is more than 200. Beyond that threshold, two curves overlap and show identical evolution. As it could be seen, as the louver angle becomes higher, the louver fins position becomes steep against flow which causes higher resistance against the FLiBe flow. Consequently, the pressure drop becomes higher and so the friction factor increases. The exact opposite can be noted too, as the angle gets smaller, the flow encounters lower resistance during the flow in particular at lower Reynolds number and it allows smoother transition over the fins.

In Fig. 11b, Colburn $j$ factor and friction factor $f$ evolution are presented, for $F_p=2.0$ mm. The overall trend is almost similar with $F_p=1.5$ mm except for the range that the curves are plotted. While the plots are in the range of 0.10 - 0.04 for $F_p=1.5$ mm, this range turns to 0.09 to 0.03 for $F_p=2.0$ mm. The evolution of the Colburn $j$-factor for all angles drowns as a cluster for the fin pitch 2 mm. The only observable plot appears for lower angles such as $L_\alpha=20°$. In this particular condition, the lowest performance is observed and plotted below the other investigated angles. As for friction factor $f$, the trend is similar to $F_p=1.5$ mm as well. The highest and lowest plots are pertaining to $L_\alpha=36°$ and 20° respectively. The difference between $L_\alpha=20°$ and 24° at higher $Re_{Lp}$ is more prominent in this case. The highest friction factor $f$ is obtained as 0.07 when the $L_\alpha=36°$ and $Re_{Lp}=100$. The major reason for the highest friction factor for $L_\alpha=36°$ is the higher resistance that molten salt encounters during the flow which escalates as the angle steeps.

Colburn $j$ factor and friction factor $f$ change for $F_p=2.5$ mm is demonstrated in Fig. 11c. Even though they have the same trend, it is becoming hard to distinguish the most efficient angle at this fin pitch since the plot cluster is more congested. The Colburn $j$ factor plots are packed in a narrower range than earlier fin pitches and the impact of the angle on the performance becomes very difficult when the space between the fins grows and flow becomes more duct oriented. Unlike Colburn, friction curves are more separated, and the difference between the curves is more obvious as the $F_p$ increases. The highest and lowest friction factor can be identified more clearly. Though the trend has not changed, the maximum friction factor $f$ is observed at around 0.43 at the $L_\alpha=36°$ when $Re_{Lp}$ is 100 while the minimum f is obtained at around 0.03 at $L_\alpha=20°$ when $Re_{Lp}$ is 500. When the overall picture is considered by taking all diagrams into account, it is seen that, all value ranges for $h$, $\Delta P$, $j$ and $f$ reduce as the $F_p$ increases.

With regard to summarized findings, thermal performance measures diminish when the fin pitch increases as a consequence of the enlargement of the uninterrupted flow region which refers to the duct-directed flow. As fin pitch becomes lower, the molten salt flow turns to louvered directed flow which enhances the interaction of the fin and FLiBe that allows improving the thermal performance. In addition, the louver angle has an important role on the pressure drop and friction either, as the angle steeps, the molten salt encounters more resistance beginning from the leading edge.

The performance of the fins is investigated by some other analogies in literature as well. One of the widely preferred criteria is the volume goodness factor ($j/f^{1/3}$) which allows considering the thermal and hydraulic performance together. The volume goodness factors evolution for all considered angles and fin pitches with respect to $Re_{Lp}$ number is presented in Fig. 12 as a whole. In the figure, the findings are presented in three individual schemes which are corresponded to the fin pitches analyzed in the study. From left to right, the diagrams are corresponding to the fin pitches of $F_p$=1.5 mm, 2.0 mm, and 2.5 mm. As it could be seen, the overall trend of the goodness factor is to descend as the $Re_{Lp}$ numbers increases. Even though every considered case has a similar tendency and obtains values in a similar range which looks like a cluster in each figure, minimum and maximum values are noticeable at each $Re_{Lp}$ number. Besides the unlikely characteristics can be identified among the curve cluster as well. From a wider aspect, for most of the considered $Re_{Lp}$ number, the curve pertaining to $F_p$=2.0 mm and $L_\alpha$=20° shows better performance while the curve belongs to $F_p$=2.5 mm and $L_\alpha$=36° shows the worst performance from the volume goodness factor aspect. The maximum goodness factor at $Re_{Lp}$ number 100, is obtained as 0.13 and for the case where $F_p$=1.5 mm and $L_\alpha$=24° while the lowest value is 0.112 for the louver fins $F_p$=2.0 mm and $L_\alpha$=36°. When the $Re_{Lp}$ number is slightly increased there is a minor change in the response of the fins for this increment, unlike the first case, $F_p$=2.0 mm and $L_\alpha$=20° shows better performance than other fin pitches and louver angles while $F_p$=2.5 mm and $L_\alpha$=36° gets the lowest value. The observed highest and lowest values for this $Re_{Lp}$ number are 0.11 and 0.09 respectively. Though the difference isn't very distinct, for the $Re_{Lp}$ number 300 and 400, the highest and lowest performance from the volume goodness aspect, are observed for the louver fins with $F_p$=2.0 mm and $L_\alpha$=20° and $F_p$=2.5 mm and $L_\alpha$=36°, respectively. Even though the order hasn't changed a lot when $Re_{Lp}$ number is 500, the lowest performance is observed for the louver fin with $F_p$=2 mm and $L_\alpha$= 32° instead. As it could be seen, the highest goodness factor is obtained as 0.13 while the lowest is observed as 0.06. As regards to outcomes, though the louver fin models with $L_\alpha$=20° don't show higher performance from both heat transfer and Colburn factor aspect, when the combined effect of Colburn $j$- and friction factors are considered, it becomes the most preferable option among the others since the pressure loss and friction factor values are considerably lower than the other cases.

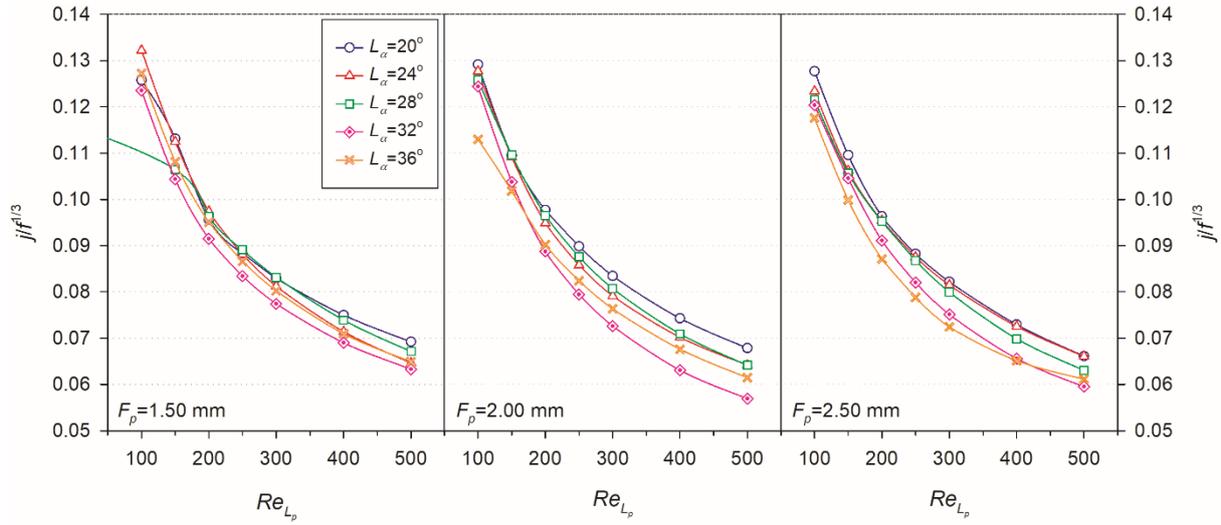

**Fig. 12.** Deviation of volume goodness factor ($j/f^{1/3}$) with respect to Reynolds number

## 4. Conclusion

Minimization of the fuel inventory in the heat exchanger of an MSR is strongly desirable. The fuel inventory should be minimized with a compact heat exchanger by increased heat transfer area and reduced volume. The most important compact heat exchanger's candidate can be louver fins over mini channeled flat tubes. The physical properties of the molten salt in the primary and secondary fluid circuits affect not only the size and cost of the heat exchanger, but also affect the pumping power, size of the piping, fuel inventory, and hence the cost of design. For the contribution to the MSR literature, the steady-state thermo-hydraulic performance of the FLiBe flowing through a CHE with multi louvered fins has been investigated with regard to the typical MSR's thermal conditions. A series of numerical tests are conducted for three different fin pitches and five different louver angles for low-velocity ranges as a preliminary work of an ongoing investigation. The steady-state velocity profiles, temperature, and pressure fields for five different Reynolds numbers are obtained and the heat transfer coefficient and pressure drop are evaluated in terms of Colburn $j$-factor and Fanning friction factor $f$, and the volume goodness factor ($j/f^{1/3}$).

The effect of louver fins defined to enhance the heat transfer was predicted. The heat transfer coefficient was improved at the expense of an increase in the pressure drop. The increased pressure drop has a positive effect on decreasing the fuel inventory. The present benchmark study is conducted for low-velocity profiles and yields to understand the appropriateness of compact heat exchangers for MSR applications by investigating the thermo-hydraulic behavior of FLiBe that flows through the louver fins in a CHX. The obtained outcomes can be highlighted as in the following:

- As the louver angle steeps the pressure at the upstream rises and while it falls at downstream, and as a consequence, the pressure drop increases as the louver angle ascends.

- As the louver angle steeps, the flow of the FLiBe becomes more louvered directed which improves the interaction between the molten salt and louver fins which enhances the thermal performance.
- As the fin pitches ascend, the flow becomes duct-directed as a result of the increasing space between the fins and this reduces the interaction between the molten salt and fins. When the interaction descends, the thermal performance of the CHX reduces as well.
- When the velocity of the molten salt flow raises, the heat transfer coefficient and pressure drop increase for all investigated fin pitches and louver angles. The highest '$h$' range is obtained for $F_p$=1.5 mm by 12000-28000 W/m$^2$K, while this range changes as 12000-25000 W/m$^2$K and 12000-24000 W/m$^2$K for the $F_p$=2 mm and $F_p$=2.5 mm respectively. A similar decrement is observed when the fin pitch impact is observed on the pressure drop. The ΔP range is 4500-0 Pa for $F_p$=1.5 mm, while it reduces to 3000-0 Pa and 2000-0 Pa for the $F_p$=2.0 mm and $F_p$=2.5 mm respectively.
- The Colburn $j$ factor and friction factor $f$ reduce as the Reynolds number increases regardless of fin pitch and louver angle. As the fin pitch increases due to the thermal performance reduction as a consequence of the lesser interaction, Colburn $j$-factor range declines as well. While it evolves in the range of 0.10 – 0.04 for the $F_p$=1.5 mm, it drops to 0.095-0.03 and 0.09-0.03 for the $F_p$=2.0 mm and $F_p$=2.5 mm respectively. The same trend is observed for the friction factor too.

The study constitutes an intersection of two fields which are Molten Salt Reactors and compact heat exchanger design with enhanced heat transfer surface area. Therefore, it is going to be helpful for the people interested in the research, the real product design, and the manufacturing of MSR's thermohydraulic systems. As it could be seen from the summarized findings, the comprehensive study of the thermohydraulic of FLiBe with the most appropriate boundary conditions to MSR's contributes to the development of more efficient compact heat exchangers for the enhancement of heat transfer from the core of the reactor to shrink volume and decrease fuel salt volume outside the core.

**Figure Captions**

**Fig. 1.** Technical drawing of the considered louvered fin
(a) 2-D cross-sectional view of *A-A* (b) 3-D appearance

**Fig. 2.** The 2-D numerical model and the boundary conditions

**Fig. 3.** (a) Mesh independency (b) Comparison of the results with the literature

**Fig. 4.** Streamlines of FLiBe flowing through the louvered fins for $F_p$=1.5 mm and $Re_{Lp}$=100
(a) $L_\alpha$=20° (b) $L_\alpha$=32°

**Fig. 5.** Pressure fields of FLiBe flow through the louvered fins for $F_p$=1.5 mm and $Re_{Lp}$=100
(a) $L_\alpha$=20° (b) $L_\alpha$=32°

**Fig. 6.** Temperature fields of FLiBe flow through the louvered fins for $F_p$=1.5 mm and $Re_{Lp}$=100
(a) $L_\alpha$=20° (b) $L_\alpha$=32°

**Fig.7.** Streamlines of FLiBe flowing through the louvered fins for $F_p$=2.5 mm and $Re_{Lp}$=400
(a) $L_\alpha$=20° (b) $L_\alpha$=32°

**Fig. 8.** Pressure fields of FLiBe flowing through the louvered fins for $F_p$=2.5 mm and $Re_{Lp}$=400
(a) $L_\alpha$=20° (b) $L_\alpha$=32°

**Fig. 9.** Temperature fields of FLiBe flowing through the louvered fins for $F_p$=2.5 mm and $Re_{Lp}$=400
(a) $L_\alpha$=20° (b) $L_\alpha$=32°

**Fig. 10.** Deviation of heat transfer coefficient (*h*) and pressure drop ($\Delta P$) with respect to frontal velocity for FLiBe ($u_{in}$) (a) $F_p$=1.5 mm (b) $F_p$=2.0 mm (c) $F_p$=1.5 mm

**Fig. 11.** Deviation of Colburn *j*- and *f* factors with respect to frontal velocity for FLiBe (a) $F_p$=1.5 mm (b) $F_p$=2.0 mm (c) $F_p$=1.5 mm

**Fig. 12.** Deviation of volume goodness factor ($j/f^{1/3}$) with respect to Reynolds number

**Table Captions**

**Table 1.** Definition of figure of merits (FOMs)

**Table 2.** Comparison of FOMs for different coolants

**Table 3.** Properties of FLiBe and Hastelloy-N